\documentclass[twocolumn,eqsecnum,preprintnumbers,nofootinbib]{revtex4}
\usepackage{amsmath,amssymb,amsfonts,times,graphicx,verbatim}
\usepackage[bookmarks=true,colorlinks,linkcolor=red,urlcolor=blue,citecolor=blue]{hyperref}
\usepackage[usenames]{color}
\usepackage{dcolumn}
\usepackage[normalem]{ulem}
\usepackage{enumerate}
\usepackage{epsfig}
\usepackage{yfonts}
\usepackage{bm}

\newcommand{\beq}{\begin{equation}}
\newcommand{\eeq}{\end{equation}}
\newcommand{\ba}{\begin{array}{ccc}}
\newcommand{\ea}{\end{array}}

\def\bea{\begin{eqnarray}}
\def\eea{\end{eqnarray}}

\begin{document}

\title{Structure of entanglement in regulated Lorentz invariant field theories}
\author{Brian Swingle}
\affiliation{Department of Physics, Harvard University, Cambridge, Massachusetts 02138}

\date{\today}
\begin{abstract}
Regulated Lorentz invariant quantum field theories satisfy an area law for the entanglement entropy $S$ of a spatial subregion in the ground state in $d>1$ spatial dimensions; nevertheless, the full density matrix contains many more than $e^{S}$ non-zero eigenvalues.  We ask how well the state of a subregion $R$ in the ground state of such a theory can be approximated when keeping only the $e^{S}$ largest eigenvalues of the reduced density matrix of $R$.  We argue that by taking the region $R$ big enough, we can always ensure that keeping roughly $e^{S}$ states leads to bounded error in trace norm even for subregions in gapless ground states.  We support these general arguments with an explicit computation of the error in a half-space geometry for a free scalar field in any dimension.  Along the way we show that the Renyi entropy of a ball in the ground state of any conformal field theory at small Renyi parameter is controlled by the conventional \textit{thermal entropy density} at low temperatures.  We also reobtain and generalize some old results relevant to DMRG on the decay of Schmidt coefficients of intervals in one dimensional ground states.  Finally, we discuss the role of the regulator, the insensitivity of our arguments to the precise ultraviolet physics, and the role of adiabatic continuity in our results.
\end{abstract}

\maketitle

\section{Introduction}
Finding useful representations of quantum states is a task of fundamental importance in the physics of quantum matter.  It is a problem that we must master to fully understand quantum matter, but it is also exceedingly practical: such understanding is almost certain to lead to powerful techniques for studying the physics of concrete and relevant physical models.  For example, the power of DMRG \cite{dmrg_orig} to efficiently solve one dimensional systems arises from an efficient representation of ground states in one dimension known as matrix product states (MPS) (see Ref. \cite{dmrg_review} for a recent review).

Entanglement plays a basic role in the search for such efficient representations.  In concrete terms, the existence of an area law \cite{arealaw1,arealaw2} for the entanglement entropy of spatial subregions in many systems suggests a generalization of matrix product states called tensor network states (or PEPS/MERA/etc.) \cite{peps,terg,mera}.  In a tensor network state the amplitude for some configuration of physical degrees of freedom is computed by contracting a large local network of tensors.  The locality of the network then guarantees that an area law for entanglement entropy is obeyed.

To make these ideas more precise, consider a bipartite quantum system divided into regions $R$ and $\bar{R}$ with the whole system in its ground state (with some unspecified local Hamiltonian).  The Schmidt decomposition of the full quantum state is
\beq
|\psi\rangle = \sum_{i = 1}^\infty \sqrt{\lambda_i} |R,i \rangle \otimes |\bar{R}, i\rangle
\eeq
in terms of the Schmidt coefficients $\lambda_i$ and orthonormal Schmidt vectors $|A,i\rangle$ ($A = R,\bar{R}$).  The reduced density matrix of $R$ is
\beq
\rho_R = \text{tr}_{\bar{R}}(|\psi\rangle \langle \psi |) = \sum_i \lambda_i |R,i \rangle \langle R, i|,
\eeq
and the entanglement entropy is
\beq
S(R) = -\text{tr}_R(\rho_R \ln{\rho_R}) = - \sum_i \lambda_i \ln{\lambda_i}.
\eeq
We say the the entanglement entropy of a region $R$ of linear size $L$ in the ground state of a local $d$-dimensional Hamiltonian satisfies an area law if the entropy scales like $S(R) \sim |\partial R| \sim L^{d-1}$.  Since $e^S$ provides a rough measure of the dimension of the support of $\rho_R$, i.e. the number of non-zero $\lambda_i$, it seems interesting to consider a class of states having the property that the density matrix of a region $R$ has support bounded by $e^{c|\partial R|}$ for all $R$.  Tensor network states provide just such a class since entanglement between $R$ and its complement $\bar{R}$ is only generated by bonds that cut the boundary of $R$.  Hence the entropy of region $R$ is bounded by $|\partial R| \log{\chi}$ where $\chi$ is the ``bond dimension" of the tensors forming the tensor network, i.e. the indices of the tensors run from $1$ to at most $\chi$.

However, we hasten to emphasize that the above discussion is heuristic at best since the physical density matrix $\rho_R$ will not in general have only $e^S$ non-zero eigenvalues.  Indeed, we expect to have many more than $e^S$ non-zero eigenvalues of $\rho_R$, so a truncation of the support of $\rho_R$ entails some error.  One precise sense in which $e^S$ is the right number of eigenvalues is in the asymptotic setting where one can show that $\rho_R^{\otimes M}$ can be well approximated by a state with support on $e^{MS}$ states provided $M$ is large enough \cite{q_shannon_code}.

In this work we are interested in the following question: what is the error in directly approximating $\rho_R$ (not $\rho_R^{\otimes n}$) by a state with only $\sim e^{S}$ non-zero eigenvalues.  We are interested in this question in part because it has bearing on the suitability of tensor network states to represent ground states of local Hamiltonians.  If, as we show, $e^S$ states suffice, then tensor network states, even with constant bond dimension, have enough degrees of freedom to represent ground states of local Hamiltonians very precisely (to small constant error in trace norm).  Such a result represents non-trivial evidence favoring the tensor network variational class beyond evidence obtained from numerical simulations of particular models.

Our questions are analogous to those asked in Refs. \cite{nature_ent,generic_ee} except that we focus on ground states of local Hamiltonians instead of generic (Haar random) states.  Since we cannot answer our question in general we focus on an interesting class of systems that can be called regulated Lorentz invariant quantum field theories.  These are lattice models or otherwise regulated models with a finite number of degrees of freedom that approximate Lorentz invariant field theories at low energy.  A review of a subclass of such models relevant for the physics of strong interactions can be found in Ref. \cite{lat_gauge_rev,subir}.  More generally, examples of these models abound in nature, including electrons at half-filling in graphene (provided we neglect or screen Coulomb interactions) and quantum magnets in the Ising universality class at their critical point.  All known topological phases in two dimensions also have representatives in this class \cite{espec_geo} (although we certainly do not claim that all topological phases are Lorentz invariant), so our tools apply there as well.

All the models we consider (in $d>1$ spatial dimensions) satisfy an area law for the entanglement entropy, $S(R) \sim |\partial R|$, even if they are gapless.  We use powerful geometrical tools that give access to the \textit{entire spectrum} of the reduced density $\rho_R$ for special subregions $R$ (balls and half-spaces) to compute errors \cite{rindler_ham,unruh_pi,holoee_deriv}.  These tools have a long history in the physics of quantum fields in curved spacetime, black holes, and quantum gravity (see Ref. \cite{susskind} for a nice introduction), and now we are putting them to work describing the structure of entanglement in certain kinds of quantum matter.  The half-space results apply to any Lorentz invariant theory while the ball results require the additional presence of conformal symmetry.  Our notion of error is usually provided by the trace norm $|| A ||_t = \text{tr}(\sqrt{A^\dagger A})$ which has a natural statistical interpretation.  Requiring that two states be close in trace norm, e.g. $||\rho_1 - \rho_2 ||_t < \epsilon$, is a stringent condition (see Appendix A for a discussion), but we will see that it is possible with rather modest resources.

We now summarize our results.  We show that the Renyi entropy $S_\alpha$, defined for general region $R$ by
\beq
S_\alpha(R) = \frac{1}{1-\alpha} \log{\left(\text{tr}( \rho_R^\alpha)\right)},
\eeq
in the special case where $R$ is a $d$-ball in the ground state of any conformal field theory (CFT) obeys an area law when $\alpha <1$ (but not not too small), and that the ground state Renyi entropy as $\alpha \rightarrow 0$ is controlled by the thermal entropy density of the CFT.  This is a perhaps surprising result, namely that ground state properties may be used to determine low temperature thermal properties.  Using a result of Ref. \cite{mps_faith} this immediately implies that bounded error can be achieved keeping only $e^{S_\alpha}$ states.  In detail, we can achieve error $\epsilon$ in trace norm by keeping $e^{S_\alpha + \delta}$ states such that $\log{\epsilon} < -\frac{1-\alpha}{\alpha}\delta$ ($0 < \alpha < 1$). Note that this bound does not directly apply to the entanglement entropy since $\alpha=1$ is excluded, but we argue that a similar bound holds even when $\alpha=1$ for the models we consider.  We show that that same is true for half-space regions in any Lorentz invariant theory, and we analyze in detail the case of a free scalar field showing explicitly that the error can be made small for a half-space.  Another product of our analysis is a bound on the decay of Schmidt coefficients even in gapless states.  As we discuss in more detail below, this enables us to recover and generalize old results about the decay of Schmidt coefficients in the context on DMRG on gapped spin chains.

Let us elaborate briefly on the issue of tensor network states.  We prove that the density matrices in question can be well approximated by their $e^{S}$ largest eigenvalues, but we do not prove that the full system has an efficient tensor network representation.  One way to rephrase this issue is to say that we consider the density matrix in the basis in which it is diagonal (thus effectively rendering the problem classical), but this basis may not be easily constructible from a tensor network state by tracing out some regions.  Our result is necessary but not sufficient if we want a constant (in system size) bond dimension $\chi$.

The remainder of this work is organized as follows.  First, we briefly review existing results related to our question.  Second, we discuss our tools and give some general arguments.  Third, we verify our general arguments with detailed error calculations for the simplest field theory: the free scalar.  We conclude with some broader discussion of our arguments and possible extensions.  In an appendix we discuss different metrics for comparing quantum states and record a pathological example illustrating the need, in principle, for our considerations in the main text.

\section{Review of existing results}
Before turning to our main arguments, we will briefly review some of the most relevant existing results regarding our main questions.  We always denote the density matrix of region $R$ by $\rho_R$ and the entropy by $S(R)$, and we define the ``entanglement Hamiltonian" $K_R$ via $\rho_R = \exp{\left( - K_R\right)}$. The Schmidt coefficients defined above are then related to ``entanglement energies" $E_i$ (spectrum of $K_R$) via $\lambda_i = e^{- E_i}$. The set of all the $E_i$ is called the ``entanglement spectrum" \cite{ent_spec}.  The area law for entanglement entropy says that the entropy $S(R) \sim |\partial R|$, a result that naturally leads us to suspect that $K_R$ may define an effective local Hamiltonian in one lower dimension.  We will see to what extent this is true below.  We will speak about efficiency loosely below, but roughly what we mean is efficiency in the formal complexity sense of polynomial effort (classical or quantum).

It was shown in Ref. \cite{hastings_1darea} that gapped phases in one dimension, in addition to obeying an area law, are actually efficiently representable as matrix product states (MPS).  The bound on the entropy is not very realistic in Ref. \cite{hastings_1darea} (many physical systems do exponentially better in terms of scaling with correlation length) and this bound was improved in Ref. \cite{improved_1darea}.  It was also argued in Ref. \cite{mps_faith} that even gapless systems may be efficiently represented if we are willing to permit a bond dimension $\chi$ for the matrices which grows with system size $n$.  Since DMRG can be run in a time that scales polynomially with $\chi$ ($\chi^3$) and since observables may similarly be measured in a time polynomial in $\chi$ we we see that even if $\chi \sim \text{poly}(n)$ the simulation is still formally efficient \cite{dmrg_review}.  Of course, in practice one is always limited in the values of $\chi$ than can be considered, but even two dimensional systems can sometimes be handled \cite{topo_kagome1,topo_kagome2}, so the method is quite powerful.

In higher dimensions it was shown in Ref. \cite{ee_gs_gap} that tensor network states exist with bond dimension $\chi \sim e^{\text{poly}(\log{n})}$ for gapped systems provided a certain bound on the number of excited states holds.  Of course, this only implies an area law up to logarithmic corrections (see also Ref. \cite{arealaw_log}) where as we believe a strict area law is actually valid for these systems.  Recall that all of our regulated Lorentz invariant models in $d>1$ obey an area law for the entropy.  Depending on how easily the network can be contracted, this result may still be called efficient.  However, in the worst case such a contraction is hard to perform, although in such a situation even a constant $\chi$ would not be efficient.  However, a similar result is not available for gapless systems in higher dimensions.  It is also physically sensible that the leading entropy scaling should not change within a phase, and this was recently rigorously shown in Ref. \cite{arealaw_stable} for gapped phases.

Despite the lack of general theorems, we do know of gapless systems or long range correlated systems that have exact tensor network representations.  A simple example is provided by a system whose wavefunction is the statistical weight of a classical spin model at classical criticality (see Ref. \cite{peps_classical} for an extensive discussion).  If the wavefunction is $\Psi[\sigma_1,....,\sigma_n] = \exp{\left(-\beta_c h[\sigma_1,....,\sigma_n]/2\right)}$ where $\beta_c$ is the critical temperature and $h$ is some classical Hamiltonian, then correlations of observables diagonal in the classical basis will be long-ranged.  Furthermore, it follows immediately that the system has a tensor network representation and hence that the area law is obeyed.

On the gapped side, we know many examples of topological phases that, in a certain topological limit, have exact tensor network representations.  These ``string-net" phases are topological states in $d=2$ without protected edge modes.  The area law is obeyed in all these phases and they have an exact and efficiently contractible tensor network representation.  Of course, we also have models of topologically trivial gapped phases that also possess exact tensor network representations, e.g. valence bond crystals, at least in a certain limit.  It is quite reasonable to suppose that for these models a constant bond dimension tensor network state is always close to the true ground state.  A final class of systems worthy of mention is the set of non-interacting fermionic topological insulators.  As shown in Ref. \cite{ti_espec_maj} in the non-interacting case (and Ref. \cite{espec_geo} in the interacting case) the entanglement spectrum of these models is quite simple, including the presence of edge modes and showing that entanglement is localized near the edge.

\section{Lorentz invariant entanglement spectra}
We now turn to our main interest, the error in truncating reduced density matrices of spatial subregions in Lorentz invariant theories.  Our tools only permit us to access certain special subregions.  For any Lorentz invariant system we have the spectrum when $R$ is a half-space \cite{rindler_ham,unruh_pi} while for any system that has conformal symmetry in addition to Lorentz invariance we also have the spectrum when $R$ is a $d$-ball \cite{Hislop:1981uh,holoee_deriv}.  As we indicated in the introduction, these results have had a profound impact on our understanding of the physics of black holes and quantum gravity.  Here we apply the same technology to answer questions of interest in the physics of quantum matter.

Consider first the half-space $x^0 = t= 0$, $ x^1 > 0$, and $x^i \in \mathbb{R}$ for $i=2,...,d$.  As shown in Refs. \cite{rindler_ham,unruh_pi,susskind,espec_geo} the density matrix for this subsystem is given by
\beq \label{rhohs}
\rho_{x^1 > 0} = \exp{\left(- 2 \pi J \right)}
\eeq
with $J$ a certain (regulated) boost generator
\beq \label{boostop}
J = \int_{x^1 > a} x^1 T^{00}.
\eeq
$T^{00}$ is a component of the stress tensor and $a$ is a UV cutoff.  Thus in the notation introduced above we have $K_{x^1 >0} = 2 \pi J$.  This expression is quite powerful because it gives us the spectrum of the reduced density matrix of a special subsystem in terms of a simple geometric problem: the spectrum of the field theory on Rindler space $ds^2 = - \rho^2 d \eta^2  + d\rho^2 + dx_i^2$ ($i=2,...,d$).  Here the Rindler coordinates $\eta$ and $\rho$ are related to the usual coordinates $x^0 = t$ and $x^1$ via
\beq
t\pm x^1 = \pm \rho e^{\pm \eta}.
\eeq
Rindler space enters precisely because $J$ is the generator of time translations in Rindler space, i.e. $J$ generates $\eta \rightarrow \eta + \delta \eta$, so the half-space density matrix is a thermal state of the time translation generator in Rindler space at a certain temperature.

This result has already been used in Ref. \cite{espec_geo} to argue for a quite general ``bulk-edge correspondence" in topological phases that connects entanglement spectra with physical edge spectra.  This correspondence states that if a physical edge has a protected edge mode then the entanglement spectrum will also contain this mode (modulo non-universal details) \cite{ent_spec,topent1}. It is easy to see how Eq. \ref{boostop} implies this result since it essentially states that the entanglement Hamiltonian looks like the physical Hamiltonian with an edge.  More properly, these two operators differ by terms localized at the edge plus terms in the bulk that don't close the bulk gap. This implies that any protected edge modes, that is modes that cannot be removed by local perturbations at the edge without closing the bulk gap, will be preserved in the entanglement Hamiltonian.  The ball result discussed below also permitted us to give an alternative computation of the topological entanglement entropy in $d=2$ \cite{espec_geo}.

If we interpret Eqs. \ref{rhohs} and \ref{boostop} locally, then they give a picture of a thermal system with a gap (or energy scale) that is growing linearly with $x^1$.  Alternatively, we may imagine that the temperature is decreasing as $1/x^1$.  The crucial point is that, for the half-space, the density matrix is the exponential of a local operator in $d$ dimensions (not $d-1$ dimensions even though the entropy scales like the boundary).  Furthermore, as shown in Ref. \cite{Hislop:1981uh,holoee_deriv}, the same is also true when $R$ is a ball and the system is conformally invariant.  In this case we must consider the problem of the field theory not on Rindler space but on hyperbolic space $H^d \times \mathbb{R}$ ($\mathbb{R}$ is time).  In both cases we know that the entanglement Hamiltonian is a local operator.

Let us use this information to estimate the entropy and recover the area law for these systems.  We focus on the half-space region. Consider a conformal field theory (CFT) at temperature $T$ in $d$ dimensions.  The thermal entropy density $s$ of this system goes like $s \sim T^d$.  This estimate is always valid at sufficiently high temperatures, that is at temperatures above all the mass scales in the system.  If the system does have a gap $m$ then at low temperature the entropy density will go like $s \sim e^{-m/T}$.  Now suppose the temperature varies as $T\sim (x^1)^{-1}$ as described above.  To compute the total entropy we integrate the entropy density over the entire half-space
\beq \label{ee_est}
S \sim \int dx \,d^{d-1}x_\perp s(T(x)).
\eeq
This integral must be cutoff at $x \sim a$ and we assume the transverse directions $x_\perp$ have size $L^{d-1}$.  The resulting entropy is
\beq
S \sim \left(\frac{L}{a}\right)^{d-1}+ ...
\eeq
which satisfies the area law even though we considered a gapless system (a CFT).  If the system has a gap then the $x$ integral will be cutoff when $x \sim 1/m$ so that only regions within $1/m$ of the boundary at $x=a$ contribute to the entropy.

We can do even better when $d=1$.  In terms of the central charge $c$ of the CFT$_{1+1}$ the entropy density is
\beq
s(T) = \frac{\pi c}{3} T.
\eeq
The temperature is given by $T(x) = 1/(2 \pi x)$ so that we find a total entropy of
\beq
S = \int_a^\xi dx s(T(x)) = \frac{c}{6} \log{\left(\xi/a\right)}
\eeq
where $\xi$ is an infrared cutoff (mass gap).  As usual, if the system is truly a CFT then $\xi$ would be replace by a finite interval length $L$.  Formally the half space has $L=\infty$ so we need an additional infrared cutoff.  Thus in the one dimensional setting we can even get the prefactor of the area law violation correct, and yet the entanglement Hamiltonian is still a local operator.

Similar estimates are also possible for the $d$-ball using the connection to hyperbolic space.  The metric of hyperbolic space may be taken to be
\beq
ds^2 = L^2 (- dt^2 + du^2 + \sinh^2{u} \,d\Omega_{d-1}^2 )
\eeq
where $L$ is the radius of the ball $R$ and $d \Omega_{d-1}^2$ is the metric on the round $d-1$ sphere.  We must consider the CFT at a finite temperature $T = 1/(2 \pi L)$ on this spacetime to compute the entanglement entropy.  Note that hyperbolic space is non-compact and hence the entropy diverges (as expected).  The large $u$ cutoff is related to the UV cutoff of the field theory by $u_c \sim \log{(L/a)}$.  The total volume of hyperbolic space is thus
\beq \label{hypvol}
V = L^d \Omega_{d-1} \int_0^{u_c} du \sinh^{d-1}{u} \sim L^d \left(\frac{L}{a}\right)^{d-1}.
\eeq
Since hyperbolic space is homogeneous we have the same entropy density everywhere, so the total entropy is
\beq
S = s V
\eeq
which again satisfies an area law since $s \sim 1/L^d$ (recall that the temperature is $\sim 1/L$).  Furthermore, we can see that when $d$ is even there is a universal subleading constant term in $S$ while when $d$ is odd there is a universal subleading logarithmic term in $S$.

Let us now use these tools to obtain and generalize some old results in one dimension.  Consider first a critical system, i.e. one with conformal invariance.  Because of the absence of curvature in one dimensional manifolds, the entropy density of the CFT on a ``hyperbolic line", that is just a finite line interval, at temperature $T$ is simply the usual result
\beq
s = \frac{\pi c}{3} T.
\eeq
Thus the spectrum of the reduced density matrix is simply identical to that of a thermal state on a finite interval of length $V = 2 L \ln{(L/a)}$ at temperature $T = (2\pi L)^{-1}$.  Using $S = s V$ we obtain
\beq
S = \frac{\pi c}{3} \frac{1}{2\pi L} (2 L \ln{(L/a)}) = \frac{c}{3} \ln{(L/a)}
\eeq
as required.  Moreover, we obtain the complete spectrum in terms of the conventional energy spectrum, a result first obtained in Ref. \cite{espec_cft_1d} using the Renyi entropy.  Our approach has the additional virtue of justifying the weak dependence of the non-universal term in $S_n$ on $n$ (effectively the temperature in our langauge).  Indeed, we see here that non-universal physics enters in the specification of the volume of the hyperbolic space and plays only a minor role.

The preceeding discussion is sufficient for studying the entropy, but if we want to ask about individual levels in the entanglement spectrum, we must specify boundary conditions at $u=u_c$, for example.  It can be explicitly checked that different boundary conditions give the same leading scaling of the entropy with the central charge, but different boundary conditions can change the detailed spectrum.  A very natural boundary condition is to demand that there be no energy flux through the surface $u=u_c$ (otherwise energy is not conserved).  For a free scalar field $\phi$ in $d=1$, the stress tensor is
\beq
T^{\mu\nu} = \partial^\mu \phi \partial^\nu \phi - \frac{1}{2} g^{\mu \nu} (\partial \phi)^2.
\eeq
The energy current is $T^{01}$, so requiring
\beq
T^{01}|_{bdy} = 0
\eeq
requires $\partial_t \phi \partial_x \phi = 0$ at the boundary.  We can satisfy this boundary condition by simply setting $\partial_x \phi = 0$.  These open boundary conditions for a scalar field yield a spectrum in agreement with the recent numerical calculations of Ref. \cite{2013arXiv1303.0741L}, so the field theory methods presented here can provide a detailed accounting of the entanglement spectrum of lattice models whose low energy limit is the field theory in question.

We can also obtain interesting results for gapped systems.  Suppose we now consider one of the above CFTs with a relevant perturbation that leads at low energies to a finite gap in the thermodynamic limit.  Because again the entanglement Hamiltonian has the form of a local Hamiltonian, it follows that states of high entanglement energy can be mapped to states of high physical energy.  Since states of high entanglement energy have correspondingly small Schmidt coefficient, any attempt to bound the decay of the Schmidt coefficients should focus on such high entanglement energy states.  The difference from the CFT case is that we should take the dimensionless volume of space to be $2 \ln{(\xi/a)}$ with $\xi$ the correlation length (recall that $L$ sets the units of all quantities).  Now the high energy density of states in a CFT in volume $V$ at energy $E$ (also made dimensionless) is
\beq
\Omega(E) = \exp{\left(\sqrt{\frac{2 \pi c}{3} V E}\right)},
\eeq
and the partition function at dimensionless temperature $T = 1/(2\pi)$ is
\beq
Z = \sum_E \Omega(E) e^{- 2\pi E}.
\eeq
This partition function is just the trace of $e^{-K}$ and hence is proportional the sum of all the Schmidt coefficients.  We can obtain a sense of how fast the Schmidt coefficients at large $E$ decay by arranging the Schmidt coefficients in non-increasing order $\lambda_1 , \lambda_2, ...$ and bounding the function
\beq
f(N) = \sum_{i=N+1}^\infty \lambda_i.
\eeq
This is nothing but
\beq
Z_>(E^\star) = \sum_{E > E^\star} \Omega(E) e^{-2\pi E}
\eeq
with $E^\star$ defined by
\beq
N = \sum_{E < E^\star} \Omega(E).
\eeq
Since $\Omega(E)$ grows as the exponential of $\sqrt{E}$, the relationship between $N$ and $E^\star$ is approximated by
\beq
N = \Omega(E^\star)
\eeq
or
\beq
E^\star = \frac{3 (\ln{N})^2}{2 \pi c V}.
\eeq
At large $E^\star$ or large $N$ we then approximate $Z_>(E^\star(N))$ by
\beq
f(N) = Z_>(E^\star(N)) \sim \exp{\left(- \frac{3 (\ln{N})^2}{2 c \ln{(\xi/a)}}\right)}.
\eeq
This is a result of relevance for the performance of DMRG first obtained for some integrable models in Ref. \cite{evalue_dist_1d}.  Here we obtain it much more easily without resorting to integrability and for any Lorentz invariant gapped state.

\section{General arguments}

We have gone to some effort to establish the intuition that these entanglement Hamiltonians may be treated locally.  Ultimately our argument is a simple extension of this observation.  In a nutshell, locality of the entanglement Hamiltonian implies that a conventional thermodynamic (large $|R|$) limit will guarantee a concentration of measure around the $e^{S}$ most probable states.  Looking, for example, at the entanglement Hamiltonian for the half-space
\beq
K = 2 \pi \int_{x^1 >a} x^1 T^{00},
\eeq
we see that the crucial point is that the linearly rising energy scale effectively cuts off the entropy in the $x^1$ direction even for gapless systems.  For a system with a gap our claim is intuitively quite reasonable, but below we will show explicitly that it is valid even for certain gapless systems.  In the case of CFTs, the situation is particularly clear when $R$ is a disk and we can use the map to the hyperbolic setting.  There we have a completely conventional thermal system, albeit in a curved space, with a uniform temperature and entropy density.

Here is yet another perspective.  Instead of asking about the full spectrum, suppose we refine this information into the Renyi entropy
\beq
S_\alpha(R) = \frac{1}{1-\alpha} \log{\left(\text{tr}( \rho_R^\alpha)\right)}.
\eeq
When $R$ is a $d$-ball we know that $\rho_R$ has the same spectrum as
\beq
Z^{-1} \exp{\left(- 2\pi L \mathcal{H}_{H^d} \right)}
\eeq
where $\mathcal{H}_{H^d}$ is the Hamiltonian on hyperbolic space.  To compute the Renyi entropy we consider the more general state
\beq
Z^{-1}_\alpha \exp{\left(- 2\pi \alpha L \mathcal{H}_{H^d} \right)}
\eeq
with $Z_\alpha = \text{tr}(\exp{\left(- 2\pi \alpha L \mathcal{H}_{H^d} \right)})$.  The limit $\alpha\rightarrow \infty$ projects onto the ground state of $\mathcal{H}_{H^d}$ while the limit $\alpha \rightarrow 0$ is a high temperature limit.  Now at very high temperatures a CFT will not detect any small local curvature since high temperature corresponds to short distances.  Thus we can use the usual flat space density of states for the CFT to estimate the free energy density.  In terms of $\alpha F_\alpha = -\log{(Z_\alpha)}$ we have
\beq
S_\alpha = \frac{\alpha F_1 - \alpha F_\alpha}{1-\alpha}.
\eeq

Now we estimate $F_\alpha$ by considering the asymptotic number of states $\Omega(E)$ of a CFT in volume $V$ at energy $E$:
\beq
\Omega(E) = \exp{\left(k (E^d V)^{1/(d+1)}\right)}.
\eeq
The partition function at high temperature is thus
\beq
Z = \sum_E \Omega(E) e^{-E/T}
\eeq
which is maximized when
\beq
\frac{1}{T} = \frac{d k }{d+1} \left(\frac{V}{E}\right)^{1/(d+1)}.
\eeq
or
\beq
E = V \left( \frac{k d T}{d+1} \right)^{d+1}
\eeq
The free energy is therefore
\begin{eqnarray}
&& F/T = - \log{Z} = - V \left( k \left(\frac{k d T}{d+1}\right)^d  - \frac{1}{T}\left( \frac{k d T}{d+1} \right)^{d+1} \right) \cr
&& = - V \frac{T^d}{d} \left(\frac{k d}{d+1}\right)^{d+1}.
\end{eqnarray}
Setting $T \sim (2 \pi L\alpha)^{-1}$ we have for $F_\alpha$
\beq
\alpha F_\alpha = V \frac{1}{d (2 \pi L \alpha)^d} \left(\frac{k d}{d+1}\right)^{d+1}.
\eeq
This result is valid at small $\alpha$ since this corresponds to high ``entanglement temperature".

Since the conventional flat space thermal entropy is
\beq
S_{th} = - \frac{\partial F}{\partial T}
\eeq
we have
\beq
S_{th} = - (d+1) (F/T).
\eeq
Now in the $\alpha \rightarrow 0$ limit we have $|\alpha F_1| \ll  |\alpha F_\alpha|$, so
\beq
S_\alpha \sim -\alpha F_\alpha
\eeq
which is indeed positive and is precisely related to the conventional thermal entropy at a particular temperature and volume.  In particular, we see that the Renyi entropy goes like (see Eq. \ref{hypvol})
\beq \label{renyilow}
S_\alpha \sim V (L\alpha)^{-d} \sim L^{d-1} \alpha^{-d}
\eeq
which still satisfies a boundary law.  This result is already quite interesting since we have shown that the Renyi entropy of a ball, as $\alpha \rightarrow 0$, is controlled by the high temperature thermal entropy in any CFT (see Ref. \cite{renyi_free} for more results for free CFTs and Ref. \cite{PhysRevB.80.115122} for results about the interacting $O(N)$ model).  There are a few subtleties in this result related to boundary conditions and topological field theories which we discuss in Appendix C.

Note that the limit $\alpha \rightarrow 0$ is ultimately singular since it eventually gives simply the Schmidt rank of the reduced density matrix.  In any regulated system with a finite number of states the growth with $\alpha$ we computed cannot continue forever.  At the very least, if there are $e^{ q |R|}$ total states in region $R$ (e.g. $q = \log{2}$ for qubits) then we must have $S_\alpha \leq q |R|$.   This trivial bound implies that our computation must fail when
\beq
\frac{1}{\alpha^d}\left(\frac{L}{a}\right)^{d-1} \sim \left(\frac{L}{a}\right)^{d}
\eeq
where $L$ is the linear size of $R$.  Thus we must at least have
\beq
\alpha > \left(\frac{a}{L}\right)^{1/d},
\eeq
but more generally we expect the CFT result to be valid provided the region is large enough and we consider $\alpha$ near one so that we are safely probing the long distance physics of the regulated field theory.

Since $\alpha < 1$ favors the smaller probability states in the tail of the distribution, our result for the Renyi entropy can also be used to establish that $e^S$ states is sufficient.  Let $\rho_M$ denote the density matrix truncated to its $M$ largest eigenvalues and let $\epsilon = ||\rho - \rho_\chi||_t$.  If we order the eigenvalues of $\rho$ so that $p_i \geq p_{i+1}$ then we have $\epsilon = \sum_{i=M+1}^{S_0} p_i$ where $S_0$ is the Schmidt rank.  In Ref. \cite{mps_faith} it is shown that the error $\epsilon$ satisfies
\beq
\log{(\epsilon)} \leq \frac{1-\alpha}{\alpha} \left(S_\alpha - \log{\left(\frac{M}{1-\alpha}\right)}\right).
\eeq
for any $0<\alpha< 1$.  Taking $M = (1-\alpha) e^{S_\alpha + \delta}$ we obtain
\beq \label{renyi_error}
\log{(\epsilon)} \leq - \frac{1-\alpha}{\alpha} \delta
\eeq
which implies that small error can be obtained by keeping roughly $e^{S_\alpha}$ states ($0 < \alpha < 1$).  Using our result for any CFT that $S_\alpha$ obeys an area law for $\alpha$ not too small, we see that again roughly $e^{S}$ states suffices.  However, if we wish to make small error in total system size $\epsilon \sim 1/\text{poly}(n)$ then $\delta$ must grow with $n$, however, if we only want small error in subsystem size, then we can still respect the $M \sim e^{S}$ scaling independent of $n$.

To conclude our general arguments, let us discuss the role of the regulator in these theories.  To have a finite entropy we must always regulate the theory, so one might worry that the regulator affects our arguments in an important way.  We argue that this is not the case because the regulator is associated with short distance physics and hence entanglement at short scales.  The regulator may affect the entanglement of nearby degrees of freedom, but its reach is spatially limited, so modifying the regulator should only modify short range entanglement.  This is explicitly visible in our geometric constructions above where the regulator is always associated with divergences near the boundary of the region.  Formally, changing the state inside a large region by adding or removing locally entangled degrees of freedom will not change the entropy of the whole region since these degrees of freedom are not entangled with the environment.  Such a transformation can be modeled by a bounded depth quantum circuit acting on the state, but clearly such a circuit cannot qualitatively affect the entanglement spectrum e.g. change its entropy scaling.  This argument is even rigorous for gapped phases up to some additional assumptions \cite{arealaw_stable}.  However, we emphasize that although we cannot rigorously prove it, in every calculation we can perform all the models we consider have the same scaling for entanglement through the entire phase whether they gapped or gapless.  Furthermore, we have a physical renormalization group understanding of this fact. Nevertheless, it is interesting to push the rigorous argument as far as we can.

One way to formalize the idea of local changes to the state within a phase is provided by quasi-adiabatic evolution \cite{quasiadiabatic}.  To begin, suppose the phase of interest is gapped and let $H(s)$ be a family of Hamiltonians within the phase.  We can produce a time dependent Hermitian operator $D(s)$ such that the ground state $|\psi(s)\rangle$ of $H(s)$ obeys $i \partial_s |\psi(s)\rangle = D(s) |\psi(s)\rangle$.  This implies that the ground state of $H(s)$ can be be obtained from the ground state at any other point, say $s=0$, by evolving for a finite time $s$ with the operator $D(s)$.  However, the crucial part of the story is that $D(s)$ can be a local operator, or at least it is local up to (almost) exponentially decaying interactions.  The length scale of the exponential decay is set by the inverse gap.  What can we say about the entropy generated by such an almost local evolution for a finite time (not growing with system size)?

In Ref. \cite{ent_cap_HAHB,arealaw_dynamics1,arealaw_dynamics2} a bound was established for the growth of entropy of a region $A$ under time evolution with a Hamiltonian of the form $H = H_A + H_B + \sum_I g_I V^I_A \otimes V_B^I$. If the $V^I$ are bounded, say $||V_A^I|| <1 $ and $|| V_B^I || < 1$, then we have
\beq
\frac{dS(A)}{dt} < \gamma \sum_I |g_I|
\eeq
with $\gamma$ some constant of order one.  Note that based on our conventions the $g_I$ all have units of energy or inverse time.  Now above we said that the interactions in $D(s)$ were exponentially localized, so let us suppose that the $g_I$ are labeled by pairs of point $I = (r_A,r_B)$ with $|g_I | < J e^{-|r_A - r_B |/\xi}$.  Then the sum over $I$ can be written as
\beq
\sum_{r_A \in A, r_B\in B} |g_I| < \sum_{r_A \in A, r_B\in B} J e^{-|r_A - r_B|/\xi},
\eeq
but this goes like the boundary $\partial A$ shared between $A$ and $B$. 

In more detail, if we assume all integrals are effectively cut off at scale $\xi$, then every term in $D(s)$ must involve no more than $\xi^d$ spins and hence can be decomposed into at most $\sim e^{\xi^d}$ product terms of the form $V_A \otimes V_B$.  Furthermore, all the terms connecting $A$ and $B$ must lie within $\xi$ of the boundary in balls of size $\xi^d$.  Thus we expect at worst a scaling like $|\partial A| e^{\xi^d} \xi^{d+1}$ for the entropy bound. Hence time evolution under a Hamiltonian like $D(s)$ for a time of order one will only add at most an area law's worth of entropy the ground state.  Thus the area law is stable within a gapped phase.  

However, it is important to understand that this is not a completely rigorous argument since we have imposed a hard cutoff at the correlation length.  Unfortunately, if we include the exponential tail then because of the crude estimate $e^{\xi^d}$ (coming from counting the number of ways to decompose an operator acting on a region of size $\xi$ into product terms) we cannot prove a bound.  Now the crude bound we used is almost certainly loose, at least for the models we consider which would not be expected to have so much entangling power.  Indeed, the maximum entanglement that could even be generated by an operator acting on a region of size $\xi$ is $\ln{e^{\xi^d}} = \xi^d$ (logarithm of the dimension of the Hilbert space).  If it were true, following Ref. \cite{2007PhRvA..76e2319B}, that when bounding the rate we could replace the dimension of the Hilbert space on which the entangling operator acts with its logarithm (as motivated by the fact that the maximum entanglement scales logarithmically), then a proof would follow.  Remarkably, independent of our considerations here, a claim to a proof of the logarithmic bound has appeared \cite{2013arXiv1304.5931V}, so the question may be settled.  It would then be interesting to extend, if possible, the results of Ref. \cite{2013arXiv1304.5931V} to the Renyi entropy as well (see below).

Going beyond the entanglement entropy, we can use the results of Ref. \cite{ent_cap_HAHB,arealaw_dynamics1,arealaw_dynamics2} with the entanglement entropy replaced by the entanglement Renyi entropy for $\alpha <1$ to argue that the Renyi entropy is also stable within a phase.  Following Ref. \cite{ent_cap_HAHB} we maximize
\beq
2 \sqrt{x(1-x)} \frac{\alpha}{1-\alpha} \frac{x^\alpha - (1-x)^\alpha}{x^\alpha + (1-x)^\alpha}
\eeq
with $x\in [0,1]$ ($x$ and $1-x$ are Schmidt coefficients of a state in a related two dimensional Hilbert space, see Refs. \cite{2001PhRvL..87m7901D,ent_cap_HAHB}).  For $1> \alpha >1/2$ this maximum exists and provides a bound on how fast $S_\alpha$ can grow under time evolution with the Hamiltonian $V_A \otimes V_B$.  Combined with the bound on the error from Ref. \cite{mps_faith} and assuming the same cutoff at the correlation length, this argument shows that our truncation results are stable within gapped phases.  There is one subtlety since such a result cannot hold as $\alpha \rightarrow 0$.  In that limit the Renyi entropy just counts the number of non-zero Schmidt coefficients which is a discontinuous quantity.  Nevertheless, the result is valid for $\alpha$ near one with the breakdown in the bound occurring for $\alpha <1/2$ since the maximization above diverges, i.e. $S_\alpha$ has no finite first time derivative.

What about gapless phases?  Clearly we cannot work below the gap since then $\xi$ would be of order the system size and nothing could be said.  However, there is another natural scale in the problem, namely the scale $\Lambda$ at which low energy physics gives way to high energy physics.  For example, in a system with emergent photons, $\Lambda$ would be the gap to states violating the low energy gauge invariance.  More generally, $\Lambda$ is the energy scale beyond which the low energy description fails and must be matched onto the high energy description.  If we now consider quasi-adiabatic evolution with the gap replaced by $\Lambda$ then one no longer takes ground states to ground states, but we can avoid mixing in states with energy greater than $\Lambda$, at least for infinitesimal times.  For finite times there will be such mixing, but we expect it can be made small so that we end up with a state that is not the ground state but has nearly the same energy density as the ground state.  These considerations are similar to those in Ref. \cite{quasiadiabatic} in the context of emergent gapless gauge fields.  The new state will by construction have no more than an area law's worth of entropy compared to the initial state.  For example, in a 1d CFT the leading logarithmic term in the entanglement entropy will not be modified by this procedure.  In this way we expect that gapless phases have the same entropy scaling throughout, although this conclusion is not completely rigorous.

Let us mention one final perspective on this issue provided by the idea of UV-IR duality (see Ref. \cite{susskind} for more discussion).  From Eq. \ref{boostop} we see that the states of the entanglement Hamiltonian of a half-space with large eigenvalue are associated not with high physical energy but with long physical distances.  Even a very low physical energy wavepacket, when placed far from the boundary, will have a large expectation value for the entanglement Hamiltonian because of the linearly rising coefficient.  Thus it is in some sense the long distance physics that is associated with states of high entanglement energy.  Such an identification makes physical sense since the question of an area law as well as the question of approximability should be depend on low energy modes, not short distance physics.  This is again explicitly visible above where the universal parts of the entanglement entropy e.g. the $\log{L}$ piece in $d=1$, are associated with long distances in Rindler space.  Since it is the effects of high entanglement energy modes that we must control to show that a low error approximation exists, it is thus quite reasonable to make such calculations within the low energy theory.

\section{Entanglement spectrum of a free relativistic scalar field}
We now make our arguments totally concrete by working out in detail the case of a half-space in the theory of a free scalar field.  We first record the spectrum of the entanglement Hamiltonian and then give a standard coding argument to address the error.

Following our discussion above, consider a $d+1$ dimensional scalar field $\phi$ in Rindler space $ds^2 = - \rho^2 d \eta^2  + d\rho^2 + dx_i^2$ ($i=2,...,d$) (see Ref. \cite{susskind} for more discussion).  The scalar wave equation is
\beq
\Box \phi - m^2 \phi = \frac{1}{\sqrt{-g}} \partial_\mu \left(\sqrt{-g} \partial^\mu \phi \right) - m^2 \phi = 0.
\eeq
Using $\sqrt{-g} = \rho$ we find
\beq
- \frac{1}{\rho^2} \partial_\eta^2 \phi + \frac{1}{\rho} \partial_\rho ( \rho \partial_\rho \phi) + \nabla_\perp^2 \phi - m^2 \phi = 0.
\eeq
Setting $\phi = \phi_\omega(\rho) e^{i k\cdot x_\perp -i \omega \eta}$ we obtain the ordinary differential equation
\beq
\omega^2 \phi_\omega + \rho \partial_\rho (\rho \partial_\rho \phi_\omega) - \rho^2 (k^2+m^2) \phi_\omega = 0.
\eeq
If we now change coordinates to the ``tortoise" coordinate $u = \log{(\rho/a)}$ then we obtain the simple equation
\beq
\partial_u^2 \phi_\omega + \omega^2 \phi_\omega - e^{2u} a^2(k^2 +m^2) \phi_\omega = 0.
\eeq
We impose two boundary conditions on this equations corresponding to the IR $u \rightarrow \infty$ and UV $u \rightarrow 0$ ($\rho \rightarrow a$) limits.  The IR boundary condition is $\phi_\omega(u \rightarrow \infty) \rightarrow 0$, and the UV boundary condition is $\phi_\omega (u = 0) = 0$.

We may get a rough idea of the spectrum by replacing these boundary conditions with box-like conditions: $\phi_\omega(0) = 0$ and $\phi_\omega(-\log{(ka)}) = 0$. The latter conditions roughly determines when the ``potential" term becomes large and forces $\phi_\omega$ to zero.  For each $k$ we then have the standard box ``energy levels"
\beq
\omega^2 = \frac{\pi^2 n^2}{(-\log{\sqrt{(ka)^2 + (ma)^2}})^2}.
\eeq
A slightly more refined estimate is provided by the semiclassical estimate
\beq
\omega \sim \frac{\pi n}{\log{\left(\frac{n}{\sqrt{(ka)^2 + (ma)^2}}\right)}}
\eeq
which reproduces our crude estimate for $n \ll ka$ and afterwards gives a slightly sublinear growth of $\omega$ with $n$.  We will use the linear spectrum for the remainder of our calculations since the results are insensitive to the subleading terms.  Although we will not need them, the exact solutions are simply Bessel functions.

As a simple check let us compute the entropy of this distribution for $d=1$ spatial dimensions (so that $k = 0$).  For a boson of frequency $\omega$ at temperature $T$ the occupation number probabilities are
\beq
p(n) = (1-e^{-\beta \omega}) e^{-\beta \omega n}
\eeq
and the entropy is
\beq
S = \sum_{n=0} - (1-e^{-\beta \omega}) e^{-\beta \omega n} \left( \log{(1-e^{-\beta \omega})} - \beta \omega n\right).
\eeq
It is convenient to compute this entropy from the heat capacity, so we study the average energy of such a collection of energy levels.  At temperature $T$ (later we set $T=(2\pi)^{-1}$ as in Eq. \ref{rhohs}) the average energy is
\beq
\langle J \rangle(T) = \sum_r \frac{\omega_r}{e^{\beta \omega_r} - 1}.
\eeq
Assuming $ma \ll 1$ we may approximate the sum by an integral as
\beq
\langle J \rangle \approx \int_0^\infty d\omega \frac{\ln{\left(\frac{1}{ma}\right)}}{\pi} \frac{\omega}{e^{\beta \omega} -1}.
\eeq
The integral may easily be computed to yield
\beq
\langle J \rangle = \frac{\ln{\left(\frac{1}{ma}\right)}}{\pi} \frac{\pi^2 T^2}{6},
\eeq
so the heat capacity and also the entropy are
\beq
C = \frac{\partial \langle J \rangle }{\partial T} = \frac{\ln{\left(\frac{1}{ma}\right)}}{\pi} \frac{\pi^2 T}{3}.
\eeq
Using $C = T \partial S/\partial T$ we see that in this case $C=S$, and putting $T=(2\pi)^{-1}$ we obtain
\beq
S = \frac{\ln{\left(\frac{1}{ma}\right)}}{6},
\eeq
which is precisely the entanglement entropy of a nearly massless scalar field with long correlation length $m^{-1}$.

We saw above that the massless scalar field in $d+1$ dimensional Rindler space ($d>1$) has a relatively simple spectrum that may be approximated as
\beq
\omega_{k r} = \frac{\pi r}{\log{\left(\frac{1}{|ka|}\right)}}
\eeq
where $k$ is the momentum in the $d-1$ transverse directions and $r=1,...$ labels the Rindler modes.  Recall that $a$ is a UV cutoff of some sort (a length) and we compactify the transverse directions in Rindler space into a torus of dimensions $L\times ... \times L$ ($L$ is the subsystem size).  The probability to observe the occupation numbers $n_{kr}$ at the Rindler temperature $2\pi$ (see above) is simply
\beq
p(\{n_{kr}\}) = Z^{-1} \exp{\left(- 2 \pi \sum_{kr} \omega_{kr} n_{kr}\right)}
\eeq
with
\beq
Z = \prod_{kr} (1 - e^{- 2\pi \omega_{kr}})^{-1}.
\eeq
Note that $p$ factorizes so that different $n_{kr}$ are uncorrelated.

From here on we give a standard coding argument adapted to our situation.  To that end, let us consider the variable
\beq
y = \left(\frac{a}{L}\right)^{d-1} \left( \log{\left(\frac{1}{p}\right)} - S \right)
\eeq
where $S$ is the entropy of $p$.  We see immediately that $\langle y \rangle =0$ since $\langle - \log{p} \rangle = S$.  We now compute the variance of $y$ as follows.  Since the $n_{kr}$ are uncorrelated we have
\beq
\langle y^2 \rangle = \left(\frac{a}{L}\right)^{2(d-1)} \sum_{kr} \left\langle \left(2 \pi \omega_{kr} n_{k r} - \frac{2 \pi \omega_{kr}}{e^{2 \pi \omega_{kr}} - 1}\right)^2 \right\rangle.
\eeq
This average may be immediately computed to yield
\beq
\langle y^2 \rangle = \left(\frac{a}{L}\right)^{2(d-1)} \sum_{kr} (2 \pi \omega_{kr})^2 \left(\frac{e^{-2\pi \omega_{kr}}}{(1-e^{-2\pi \omega_{kr}})^2}\right).
\eeq
We want an upper bound on the variance of $y$ so using
\beq
\frac{1}{1-e^{-\alpha r}} \leq \frac{1}{1-e^{-\alpha}}
\eeq
(valid for $\alpha$ positive) we have
\beq
\langle y^2 \rangle \leq \left(\frac{a}{L}\right)^{2(d-1)} \sum_{k} (2\pi \omega_k )^2 \frac{1}{(1-e^{-2\pi \omega_k})^2} \sum_r r^2 e^{- 2 \pi \omega_k r}
\eeq
with $\omega_k = \omega_{kr}/r$ independent of $r$.

Doing the sum over $r$ finally gives
\begin{eqnarray}
&& \langle y^2 \rangle \leq \left(\frac{a}{L}\right)^{2(d-1)} \sum_{k} (2\pi \omega_k )^2 \frac{1}{(1-e^{-2\pi \omega_k})^2} \cr
&& \times \left( \frac{e^{2\pi \omega_k}(1+e^{2\pi \omega_k})}{(e^{2\pi \omega_k}-1)^3}\right).
\end{eqnarray}
We now convert the sum over $k$ to an integral with the upper limit $|k| \sim 1/a$ (the UV cutoff) to find
\begin{eqnarray}
&& \langle y^2 \rangle \leq \left(\frac{a}{L}\right)^{(d-1)} \frac{a^{d-1} \Omega_{d-2}}{(2\pi)^{d-1}} \int_0^{1/a} dk k^{d-2} (2\pi \omega_k )^2 \cr
&& \times \left( \frac{e^{6\pi \omega_k}(1+e^{2\pi \omega_k})}{(e^{2\pi \omega_k}-1)^5}\right)
\end{eqnarray}
where $\Omega_{d-2}$ is the volume of the $d-2$ sphere.  Let us now analyze the limits of this expression.  First, since $\omega_k \rightarrow \infty$ when $ka \rightarrow 1$ and since the large $\omega$ limit of integrand goes like $e^{-4\pi \omega}$ we see that the upper limit of the integral strongly converges to zero.  Second, the integrand goes like $\omega^{-3}$ at small $\omega$ which occurs when $k \rightarrow 0$.  Near the $k=0$ limit we must thus consider an integral of the form
\beq
\int_0 dk k^{d-2} (-\log{ka})^3
\eeq
which still converges despite the weak logarithmic singularity.  Thus at the end of the day the whole integral is bounded by some constant $C'$ independent of $L$ and $a$:
\beq
\langle y^2\rangle \leq C' \left(\frac{a}{L}\right)^{(d-1)}.
\eeq

Using this result along with the observation that higher moments of $y$ fall off even faster with $L$ permits us to approximate the probability distribution for $y$ as
\beq
p(y) = \frac{\exp{\left(-\frac{y^2}{2\langle y^2 \rangle}\right)}}{\sqrt{2 \pi \langle y^2 \rangle}}.
\eeq
If we now want the probability that $|y| > \delta$ to be less than $\epsilon$ we see that asymptotically we need
\beq
\frac{\delta^2}{\langle y^2 \rangle } \sim \log{\left(\frac{1}{\epsilon}\right)}.
\eeq
Thus we conclude that
\beq
\left(\frac{L}{a}\right)^{(d-1)} > \frac{C'}{\delta^2} \log{\left(\frac{1}{\epsilon}\right)}
\eeq
to have small error $\epsilon$.  Note that if the variance is parametrically smaller than our estimate then this condition will be achieved even more quickly as a function of $L$.

How many states have $|y| < \delta $? The probability of all these states is greater than $(1-\epsilon)$ by assumption.  Hence we have
\beq
1 \geq \sum_{|y| < \delta} p(\{n_{kr}\}) \geq 1-\epsilon,
\eeq
but we also know from the definition of $y$ that
\beq
e^{-S + (L/a)^{d-1} \delta} \geq p \geq e^{-S - (L/a)^{d-1} \delta}.
\eeq
Hence the number $N(\epsilon, \delta)$ of states with $|y| < \delta$ is bounded by
\begin{eqnarray}
&& 1 \geq \sum_{|y| < \delta} p(\{n_{kr}\}) \geq \sum_{|y|< \delta} e^{-S - (L/a)^{d-1} \delta} \cr
&& = e^{-S - (L/a)^{d-1} \delta} N(\epsilon,\delta)
\end{eqnarray}
and
\begin{eqnarray}
&& 1-\epsilon \leq \sum_{|y| < \delta} p(\{n_{kr}\}) \leq \sum_{|y| < \delta} e^{-S + (L/a)^{d-1} \delta} \cr
&& = e^{-S + (L/a)^{d-1} \delta} N(\epsilon,\delta).
\end{eqnarray}
Combining these inequalities we have
\beq
e^{S + (L/a)^{d-1} \delta} \geq N \geq (1-\epsilon) e^{S - (L/a)^{d-1} \delta}.
\eeq
What this finally tells us is that for sufficiently large $L$ we can always obtain a good approximation to the density matrix of the massless scalar in a certain region of linear size $L$ provided we keep roughly $e^{S}$ states.

We have not been very careful about the zero mode in our calculation above, indeed, its presence depends on boundary conditions anyway.  We can always remove this zero mode by adding a small regulator mass, and since the contribution of zero mode to the variance of $y$ is logarithmic, a very small mass (exponentially small in subsystem size) suffices to ensure that the zero mode can be regulated.  In most cases such a small mass will have no effect on the physics, although one place where the zero mode does play an interesting role is the case of symmetry broken phases (see Ref. \cite{ee_ssb}), but this is beyond our scope here.

\section{Discussion}

We have shown that for a wide variety of systems obeying the area law, it suffices to keep $e^S$ states in the reduced density matrix to approximate said density matrix to constant error in trace norm.  Such a result is necessary but not sufficient if we want to prove that a tensor network with bounded bond dimension $\chi$ can represent the pure ground state of the whole system.  To demonstrate these results, we have made heavy use of geometric technology available for systems which display Lorentz and conformal invariance at low energies.  Indeed, our whole argument is ultimately based on a renormalization group picture of the structure of entanglement in local quantum systems.  Our results do invite some speculation about the ability of tensor network states to represent ground states.  For example, to the extent that we are interested only in universal quantities, tensor network states and especially the MERA \cite{mera} (because of its renormalization group structure) should be in the same phase even with bounded $\chi$.  However, depending on how small we wish to make the error, e.g. $\epsilon \sim 1/\text{poly}(n)$, we will need to make $\chi$ scale with $n$ (the system size).  For example, Eq. \ref{renyi_error} shows that $\delta \sim \log{n}$ will achieve this error bound, but this implies that no bounded $\chi$ will suffice as $n\rightarrow \infty$.  It would be very interesting to provide further support for these speculations, perhaps by looking at the eigenvectors of the entanglement Hamiltonian.

Our results also shed light on the physical meaning of the entanglement entropy, i.e. that it really does control the effective Schmidt rank in some cases.  It would be interesting to consider ``one-shot" versions of quantum information protocols, e.g. entanglement distillation in light of our results here.  For example, the max/min-entropies, which reduce in our case to various Renyi entropies, have an operational interpretation \cite{maxmin_ent_operation}.  The single-copy entanglement is the amount of entanglement deterministically distillable from the single copy of a state and is determined by the $\alpha \rightarrow \infty$ limit of the Renyi entropy.  Our results here suggest that we might be able to reliably distill more than $S_\infty$ singlets from a quantum state of the type considered above.  Finally, the mutual information also has a nice operational meaning in the asymptotic setting where it represents the total amount of classical and quantum correlation between two regions.  It would be interesting if this result could be strengthened to the non-asymptotic setting for regulated Lorentz invariant field theories.

It is also interesting to observe that the argument around Eq. \ref{ee_est} estimating the entanglement entropy by integrating an appropriate entropy density can be generalized, at a heuristic level, to many other systems.  Below we always suppress the dimensionful constant that converts temperature to some power of length, e.g. the Fermi velocity.  Consider, for example, a scale invariant theory with dynamical exponent $z\neq 1$ relating energies to momenta $\omega \sim k^z$.  The thermal entropy density in such a theory is $s\sim T^{d/z}$ but in keeping with the scaling structure we must also take $T \sim 1/x^z$ instead of $T \sim 1/x$.  Eq. \ref{ee_est} then leads to a very similar estimate for the entanglement entropy as in the relativistic case.  On the other hand, consider a Fermi liquid with thermal entropy density $s\sim k_F^{d-1} T$ and with $T \sim 1/x$ (since $z=1$ effectively).  Eq. \ref{ee_est} now gives an entanglement entropy of roughly $(k_F L)^{d-1} \log{L}$ in rough agreement with the known result (we used $L$ as an IR cutoff).  Finally, even non-Fermi liquids give the same logarithmic behavior since although the entropy density is now $s\sim T^{1/z}$ we have again $T \sim 1/x^z$ which yields the same logarithmic estimate.  Of course, this is only a heuristic since, for example, some $z=2$ critical theories, like fermions at the bottom of a quadratic band, have a factorized ground state and hence no entanglement.

In conclusion, even though we have not shown that efficient tensor network representations exist, our work does support the general notion that these states are in the right corner of Hilbert space to describe ground states of local Hamiltonians.  Furthermore, it is important to perform first principles calculations of the type described here since they provide complementary information to numerical studies of particular lattice models.  Of course, even if such tensor networks exist, it may still be difficult to work with them, for example, quite apart from complexity considerations, the (polynomial) scaling of MERA computations with $\chi$ is unfortunately quite fast.  Still, the future looks quite bright for these tensor network methods.

\textit{Acknowledgements.}  We thank J. Sau, M. Barkeshli, J. McGreevy, N. Iqbal, M. Coudron, and P. Hayden for helpful discussions related to this work.  BGS is supported by a Simons Fellowship through Harvard University.

\appendix

\section{Comparing quantum states}

Here we briefly review the metrics we consider for comparing two quantum states.  There are a number of sensible ways we might compare two states, some more stringent than others.  Let us first discuss a metric that focuses on universal physics and then turn to our more stringent quantum information criterion.

Let $H(g,n)$ be a family of local Hamiltonians with ground state $|\psi(g,n)\rangle$ and energy $E(g,n)$.  The parameter $g$ is interpreted as a coupling constant which may varied to produce different quantum phases of matter while the parameter $n$ is a measure of the system size, the total number of sites.  The thermodynamic limit is obtained by sending $n$ to infinity.  The states $|\psi(g,n\rightarrow \infty)\rangle$ are said to be in the same phase provided physical observables are smooth functions of $g$.  Often one focuses on analyticity of the ground state energy as the definition of a phase, however, this is not a particularly physically useful distinction.

A gapped phase is one where a gap in the spectrum of $H(g,n)$ persists (does not decay with $n$) as $n\rightarrow \infty$ while a gapless phase is one where this does not occur.  Typically, the gap of $H(g,n)$ would go to zero as some power of $1/n$ in a gapless phase.  In the cases of interest to us, if $d$ is the spatial dimension, then the gap will close as $n^{-1/d}$.  Of course, these definitions are only really sketches and complications do arise.  For example, we may want to restrict $n$ to be even in a symmetric gapped phase of spin-1/2 moments so that the ground state can be a singlet.  Otherwise, the extra spin-1/2 moment will lead to a degeneracy that is not accounted for in our crude definition above.

We emphasize, however, that the limit $n\rightarrow \infty$ is an artificial one.  Real physical systems always have a finite effective size, that is a finite $n$.  Sometimes the effects of finite $n$ are quite important e.g. for fluctuations in mesoscopic systems or surface effects, and sometimes they matter little as in the bulk of gapped phases.

Let us now consider two different values of $g$ in the same phase.  What is the overlap between $|\psi(g_1,n)\rangle$ and $|\psi(g_2,n)\rangle$?  In general we would expect $\langle \psi(g_1) | \psi(g_2) \rangle \sim e^{-n}$.  In terms of the trace norm we expect $|| \rho_1 - \rho_2 ||_t \sim 2 - e^{-n}$ and since $2$ is the largest value possible for normalized states we see that, statistically, these two states are maximally distinguishable.  Nevertheless, both states represent the same phase of matter.  What does this mean?

Here are two examples.  If the system in question was a two dimensional gapped topological phase then the topological entanglement entropy would be the same in $|\psi_1\rangle $ and $|\psi_2\rangle$ \cite{topent1,topent2}.  Similarly, if the system in question was a gapless phase or critical point and $g$ was an irrelevant perturbation then the long distance decay of correlations, that is operator scaling dimensions, would be the same whether they were measured in $|\psi_1\rangle $ or $|\psi_2\rangle$ \cite{eeqft,geo_ent,vidal_crit}.  Of course, we are simply reviewing the completely standard notion of universality.  Our point is merely that two states need not have high overlap or be statistically similar in order to described the same universal physics.

That being said, sometimes we want a more stringent notion of equivalent states e.g. for quantum computational tasks.  A more stringent criterion than the universality notion above is provided by demanding small error in trace norm between two states.  For example, consider the two ferromagnetic states
\beq
|\psi_1 \rangle = |\uparrow \rangle^{\otimes n}
\eeq
and
\beq
|\psi_2 \rangle = (\sqrt{1-a^2}|\uparrow \rangle + a|\downarrow \rangle)^{\otimes n}
\eeq
with $a \ll 1$.  Their overlap is $\sqrt{1-a^2}^n$ and hence their difference in trace norm is
\beq
||\rho_1 - \rho_2 ||_t = 2\sqrt{1 - (1-a^2)^n}.
\eeq
Requiring that this be small ($ < \epsilon$) forces us to choose
\beq
1 - \epsilon^2/4 < (1-a^2)^n
\eeq
or roughly
\beq
n a^2 < \epsilon^2 .
\eeq
In other words, the states on each site must be close to identical with an error going like $\epsilon/\sqrt{n}$.  Such a condition leads to nearly identical extensive quantities, not merely close results for intensive quantities and universal observables.  For example, if we measure the total $S_z$ then we find that these two states differ in spin by an amount of order $na^2 \sim \epsilon^2 $.

A constant $\epsilon$ (independent of $n$) already implies that the states being compared have non-vanishing overlap even in the thermodynamic limit.  Indeed, $|| \rho_1 - \rho_2 ||_t = 2 \sqrt{1 - |\langle \psi_1 | \psi_2 \rangle |^2} = \epsilon$ clearly gives a finite overlap.  Such a condition should suffice to argue that the two states are in the same phase in the thermodynamic limit.  For example, a change $\delta g$ in the coupling constant can generally be expected to degrade the overlap like $\langle \psi(g) | \psi(g+\delta g)\rangle \sim e^{- n \delta g}$, so to achieve a constant overlap as $n \rightarrow \infty$ we must send $\delta g \rightarrow \epsilon^2/n$. Thus we expect in general that finite overlap corresponds to the same phase with regard to universal physics.  However, we may be interested in an even stricter notion of approximation where the error decreases with $n$ e.g. $\epsilon \sim 1/\text{poly}(n)$, and in this case we are really setting the two states equal to within the maximum resolution achievable in principle on a quantum computer.

For the purposes of extracting universal physics the trace norm condition is unnecessarily demanding.  Thus we may hope that different numerical techniques, especially those like tensor network states whose efficiency is based on keeping a limited number states, have a reasonable chance of accessing universal information in a quantum phase.  On the other hand, if it were possible that even small trace norm error could be achieved using such a tensor network state, then clearly our hope would be fulfilled (although there would remain the question of efficiently finding such representations).

Ultimately, our point of view is the following.  The trace norm condition is very stringent, but if this condition can be satisfied with modest resources in a tensor network setting, then surely tensor networks can reproduce the universal physics of local Hamiltonians.  In this way we can provide non-trivial evidence, evidence that complements existing numerical studies, in favor of tensor network descriptions for a wide range of phases and phase transitions.

\section{A pathological example}
To illustrate why, in principle, we need to worry about exactly how many states need to be kept, we will now discuss a very pathological example.  In this section we discuss only a classical probability distribution and its associated Renyi entropy.  Since the quantum density matrix can always be diagonalized, we may simply think of this probability distribution as the diagonal elements of some density matrix in the appropriate basis.

The many-body system will be bit strings $x_1 ... x_n$ of length $n$.  The probabilities are
\beq
p(0...0) = (1-q) + \frac{q}{2^n}
\eeq
and
\beq
p(else) = \frac{q}{2^n}.
\eeq
In simple terms, with probability $(1-q)$ we take the string $0...0$ while with probability $q$ we take a random $n$-bit string (including potentially $0...0$). Clearly we have $\sum p =1$ since
\beq
(1-q) + \frac{q}{2^n} + \frac{(2^n-1)q}{2^n} = 1.
\eeq

This probability distribution can be understood as a thermal state for a very strange Hamiltonian.  Let $x_i$ be the value of the $i$th bit and consider the Hamiltonian $H_{bad} = - \prod_i (1-x_i)$.  The thermal state at inverse temperature $\beta$ is $\rho = e^{-\beta H_{bad}}/Z$ with
\beq
Z = e^{\beta} + 2^n - 1.
\eeq
Choosing $\beta$ so that
\beq
\frac{e^\beta}{e^\beta + 2^n - 1} = (1-q) + \frac{q}{2^n}
\eeq
we obtain the probability distribution above.  Note that $H_{bad}$ is very non-local and the inverse temperature satisfies $\beta \sim n$.

Now we compute the Renyi entropy $S_\alpha$ which is
\beq
S_\alpha = \frac{1}{1-\alpha} \ln{\left(\left((1-q) + \frac{q}{2^n}\right)^\alpha + (2^n - 1)\left(\frac{q}{2^n}\right)^\alpha\right)}.
\eeq
This expression turns out to be quite delicate in the large $n$ limit.  For $\alpha <1$ then $n\rightarrow \infty$ we have
\beq
S_\alpha = n \ln{(2)} + \mathcal{O}(1).
\eeq
For $\alpha=1$ then $n\rightarrow \infty$ we have
\beq
S_1 = q n \ln{(2)}.
\eeq
Finally, for $\alpha>1$ then $n\rightarrow \infty$ we find
\beq
S_\alpha = \mathcal{O}(1)
\eeq
All this shows that $S_\alpha/n$ is badly discontinuous at large $n$, indeed, $S_\alpha$ for $\alpha >1$ is not even extensive.

Now we evalute the trace norm criterion above.  Restricting to the largest $\chi = e^{S_1}$ states we find that
\beq
||p-p_\chi||_1 = \frac{(2^n-\chi) q}{2^n} \sim q
\eeq
since $2^{n} \gg 2^{nq}$.  Thus by taking $q$ close to one we obtain an arbitrarily bad approximation.  On the other hard, consistent with the theorem in Ref. \cite{mps_faith}, the Renyi entropy for any $\alpha <1$ does tell us how many states we need for a good ``approximation", but this number is simply all the states!  And again, the Renyi entropy $\alpha >1$ is not even extensive.

We emphasize that this is a highly artificial and contrived example.  Nevertheless, it indicates that in principle there is something to prove.  It is especially intriguing to note that while the Renyi entropy for $\alpha > 1$ totally fails to capture the right number of states (even qualitatively) in the above model, we will show that these entropies are qualitatively similar to the entanglement entropy $S_1$ and even to $S_{\alpha <1}$ in our models.  This observation is important in the context of numerical work since since the Renyi entropy is most easily computed when $\alpha = 2,3,...$.

\section{The thermal problem on hyperbolic space}

In this section we discuss some of the subtleties associated with our result relating the thermal entropy density of a CFT to the ground state Renyi entropy of a ball.  We obtained this relation using the mapping of the entanglement spectrum of the $d$-ball to the spectrum of a Hamiltonian of the CFT on hyperbolic space.  Because hyperbolic space is homogeneous, it is very natural to conjecture that the entropy of a thermal system on hyperbolic space is simply the volume of hyperbolic space times the entropy density.  Unfortunately, the volume of hyperbolic space is infinite and requires regularization, and as is often the case, this regularization breaks some of the nice symmetry properties of hyperbolic space.  In particular, a boundary is introduced and boundary conditions should be imposed.  Often these conditions do not affect the universal properties of the entropy provided they obey certain reasonable properties, e.g. preserve gauge invariance. This is because the size of the boundary at the cutoff $u_c = \ln{(L/\epsilon)}$, using the metric
\beq
ds^2 = L^2 (-dt^2 + du^2 + \sinh^2{u} d\Omega_{d-1}^2 ),
\eeq
is
\beq
vol(u=u_c) = \Omega_{d-1} L^{d-1} \left(\frac{L}{2 \epsilon}\right)^{d-1}
\eeq
to leading order in $L/\epsilon$.  Thus this boundary size is UV divergent and hence naively contributes only to non-universal quantities.

Nevertheless, the above argument is a little too quick as the following considerations show.  Consider the special case of a CFT which is also a topological field theory.  Not only does such a system not care about the size of the metric, it doesn't care about the metric at all.  Let us think about $d=2$ topological states.  How can we recover the fact that there is a universal topological entropy that doesn't depend on Renyi index (in contrast to the $\alpha$ dependence in Eq. \ref{renyilow})?  A simple example is provided by $U(1)$ Chern-Simons theory.  This theory is a conformal field theory with zero correlation length in the limit of infinite gauge field mass, nevertheless, the theory possesses gapless edge states on any manifold with boundary no matter how large the bulk gauge field mass is.  We already indicated above that hyperbolic space, when regulated, has a boundary. Thus although the bulk is inert, the entanglement spectrum (spectrum on hyperbolic space) always has edge states.  It is these edge states which lead to topological entanglement entropy.

In fact, we may explicitly show this as follows \cite{topent1}.  The partition function of such $1+1$ dimensional edge states with Hamiltonian $H$ and temperature $\beta^{-1}$ is given by
\beq
Z_{a} = \text{tr}_a(e^{-\beta H}),
\eeq
where $a$ denotes the topological sector (boundary condition); we are interested in the identity sector $a=1$.  A modular transformation $\tau \rightarrow -1/\tau$ ($\tau_x$, $\tau_y$ are the two sides of the space-imaginary-time torus) exchanges space and time and acts on $Z_a$ as
\beq
Z_a(\tau) = S_a^b \tilde{Z}_b(-1/\tau).
\eeq
We already saw above that the temperature is fixed at $\beta = 2\pi L$, but the size of the boundary is large $\sim L (L/\epsilon)$.  Such a modular transformation is useful because all the modularly transformed terms $\tilde{Z}$ have length $\sim L$ and inverse temperature $L^2/\epsilon$ and hence are exponentially close to the ground state.  Thus we need only keep track of the Casimir energy of the identity sector plus the modular S-matrix to compute the partition function of interest.  At the end of the day the universal properties are controlled by the modular $S$-matrix $S_a^b$ and in particular we find an entropy of $\ln{S_1^1} = - \ln{\mathcal{D}}$ in terms of the total quantum dimension $\mathcal{D}$.  This was the line of argument followed in Ref. \cite{espec_geo} to establish the existence of edge states in the entanglement spectrum.

Finally, we see that the other important property of the boundary contributions is that they are expected to scale as a lower power of ``entanglement temperature," e.g. $1/\alpha$.  Hence in the limit of $\alpha \rightarrow 0$, where we claimed that the Renyi entropy was controlled by the conventional thermal entropy density, the boundary contributions are parametrically supressed and our bulk argument is valid.  More generally, there will be subleading in $1/\alpha$ corrections to this statement; these corrections are amply demonstrated by the case of topological theories which properly have zero thermal entropy (in the limit of infinite mass as required to treat them like CFTs) but do have a small $\alpha$ independent Renyi entropy.  Thus our statement applies to the $1/\alpha^d$ term in the Renyi entropy which is always controlled by the conventional thermal entropy even when this entropy is zero.

\bibliography{expS_lorentz}

\end{document}